\documentclass[%
 reprint,
 amsmath,amssymb,
 aps,
]{revtex4-2}

\usepackage{graphicx}   % Include figure files
\usepackage{dcolumn}    % Align table columns on decimal point
\usepackage{bm}

\begin{document}

\preprint{APS/123-QED}

\title{Density anomaly in water-alcohol mixtures: minimum model for structure makers and  breakers}
%\thanks{}

\author{Marco A. Habitzreuter}
\author{Marcia C. Barbosa}
\email{marcia.barbosa@ufrgs.br}
\affiliation{
Instituto de Física - Universidade Federal do Rio Grande do Sul\\ Porto Alegre, 91501-970, Rio Grande do Sul, Brazil.
}

\begin{abstract}
We modeled the change in the temperature of maximum density (TMD) of a water-like solvent when small amounts of solute are added to the mixture. The solvent is modeled as a two length scales potential, which is known to exhibit water-like characteristic anomalies, while the solute is chosen to have an attractive interaction with the solvent which is tuned from small to large attractive potential. We show that if the solute exhibits high attraction with the solvent it behaves as a structure maker and the TMD increases with the addition of solute, while if the solute shows a low attraction with the solvent the TMD decreases, with the solute behaving as a structure breaker.
\end{abstract}

\maketitle
%%%%%%%%%%%%%%%%%%%%%%%%%%%%%% INTRODUCTION %%%%%%%%%%%%%%%%%%%%%%%%%%
\section{Introduction}
%%%%%%%%%%%%%%%%%%%%%%%%%%%%%% INTRODUCTION %%%%%%%%%%%%%%%%%%%%%%%%%%

There are a variety of thermodynamic, dynamic and structural properties in which water behaves differently when compared with other materials. For certain pressure and temperature ranges, the density of water increases with temperature, exhibiting a temperature of maximum density (TMD), the isothermal compressibility~\cite{kanno1979}
and the heat capacity~\cite{franks2000water} increase as the temperature is decreased and  the diffusion coefficient of supercooled water increases with pressure at constant temperature~\cite{angell1976,prielmeier1987,netz2001jcp}. Even though the molecule was widely studied, some thermodynamic, dynamic and structural anomalous behavior of water are still not completely understood.

In the 70s, core-softened (CS) potentials were introduced~\cite{hemmer1972coresoftened}. They exhibit a hard core and an attenuated region, like a ramp or a step, and display a region of pressure and temperature with an anomalous behavior in density~\cite{debenedetti1991coresoftened_tmd}, diffusion and isothermal compressibility~\cite{alan2008,scala2001}. The anomalies in these simple models originate from the competition between two length scales~\cite{alan2008,alan2006,furlan2017} which resemble the water-water interactions. The shoulder, which is the shorter scale, represents the nonbonding hydrogen interactions and the longer scale, the well, represents the bonding interactions between different water molecules. The first is more relevant at high pressures and the later at lower temperatures. Although atomistic models to describe water are widely available~\cite{berendsen1987,abascal2005,mahoney2000,russo2014,holten2014,vega2009}, CS potentials are simpler and they can be useful in the understanding of the fundamental mechanism behind water or of water-like behavior.

Water can also exhibit an unusual behavior when mixed with other substances. The excess volume of water with alcohols~\cite{patel1985,sha-2016,ott-1993}, with alkanolamines~\cite{maham-1994,stec-2014} or with hydrophilic ionic liquids is negative while the excess volume of water with hydrophobic ionic liquids is positive~\cite{maham-1994,stec-2014}.
The excess enthalpy of water with  methanol~\cite{tomaszkiewicz1986,lama1965} and ethanol~\cite{lama1965,ott1986ethanol,ott1987ethanol} for most temperatures is negative, but it is positive for larger alcohols~\cite{marongiu-1994}. The excess specific heat is positive for water with ethanol~\cite{ott1987ethanol} and tert-butanol~\cite{blacet1931,visser1977}.

The addition of small amounts of solute also impacts the water TMD. Materials which increase the temperature of maximum density of water when mixed with water such as isopropanol, tert-butanol, sec-butanol, 2-butanol~\cite{wada1962}, ethyl and n-propyl alcohols~\cite{mchutchison1926anomaly} are called ``structure-makers''. Solutes which decrease the TMD are called ``structure-breakers'', such as ethylene glycol, glycerin and phenol~\cite{wada1962}. They latter tend to weaken the hydrogen bond structure and less temperature is necessary to reach the minimum volume, while the former strengthen the network.

In order to understand the mechanism behind the behavior of structure makers and breakers a number of effective models were developed. For instance, a lattice model shows that the addition of a hard sphere solute in a water-like solvent, disrupts the water hydrogen bonds, decreasing the TMD~\cite{szortyka2012}. If the solute-solvent becomes attractive, beyond a threshold attraction, the solute becomes structure maker~\cite{girardi2015}.

Here we propose that the minimum ingredient for a structure-maker is a solute-solvent threshold attraction. We test this hypothesis by computing the TMD using both a one dimensional exact solution and three dimensional Molecular Dynamics simulations which employ a simple model: a mixture of a two length scale solvent and a solute which exhibits a short range attractive interaction with the solvent. The solute itself is also a simple monomer, in contrast with more sophisticated alcohol models~\cite{murilo2020,urbic2015}. The two dimensional system, even though theoretically interesting, will not be analyzed here since we are looking for an explanation of a phenomena that appears in three dimensions.

The remainder of the paper is structured as follows. In Section~\ref{sec:model}, the one dimensional model and the analytic solution are presented. In Section~\ref{sec:3d}, the three dimensional system is presented and analyzed by Molecular Dynamics simulations. Our concluding remarks are in Section~\ref{sec:conclusion}.
%%%%%%%%%%%%%%%%%%%%%%%% 1D %%%%%%%%%%%%%%%%%%%%%%%%
\section{One Dimensional System}
\label{sec:model}
%%%%%%%%%%%%%%%%%%%%%%%% 1D %%%%%%%%%%%%%%%%%%%%%%%%
%%%%%%%%%%%%%%%%%%%%%%%%%%%%%% 1d - MODEL %%%%%%%%%%%%%%%%%%%%%%%%%%
\subsection{Model and Analytic Method}
%%%%%%%%%%%%%%%%%%%%%%%%%%%%%% 1d - MODEL%%%%%%%%%%%%%%%%%%%%%%%%%%
The system is composed of a solvent, called $A$, and a solute denoted by $B$. $N_{A}$ and $N_{B}$ represent the number of particles of each type and we denote $N$ the total number of particles. The quantities $N_{AA}$, $N_{AB}$ and $N_{BB}$ represent the number of first-neighbor interactions of types AA, AB and BB respectively. The interaction potential between a particle of type $i$ and a particle of type $j$ is represented by $V_{ij}(r)$. For simplicity, we assume $A$ and $B$ have the same mass $m$.

In one dimension, the partition function in the Isothermal-Isobaric Ensemble for this system is~\cite{prigogine1957molecular} (for details see Appendix~\ref{sec:partitionfunction:appendix}):
%%%%%%%%%%%%%%%%%%%%%%%% EQUATION %%%%%%%%%%%%%%%%%%%%%%%%
\begin{equation}
    Y(\beta, P, N) = \frac{1}{\Lambda ^N} \frac{N_A! N_B! \varphi_{AA}^{N_{AA}} \varphi_{AB}^{N_{AB}} \varphi_{BB}^{N_{BB}}}{N_{AA}!N_{BB}! \left[\left(\frac{N_{AB}}{2}\right)!\right]^2},
\end{equation}
%%%%%%%%%%%%%%%%%%%%%%%% EQUATION %%%%%%%%%%%%%%%%%%%%%%%%
with
%%%%%%%%%%%%%%%%%%%%%%%% EQUATION %%%%%%%%%%%%%%%%%%%%%%%%
\begin{align*}
    \Lambda &\equiv \left(\frac{\beta h^2}{2\pi m}\right)^{1/2}, \\
    \varphi_{ij} &\equiv \int_{0}^{\infty} e^{-\beta[V_{ij}(r)+Pr]}dr,
\end{align*}
%%%%%%%%%%%%%%%%%%%%%%%% EQUATION %%%%%%%%%%%%%%%%%%%%%%%%
where $\beta=1/(k_B T)$ and $h$ is the Planck constant.

The Gibbs Free Energy for this mixture is (for details see Appendix~\ref{sec:gibbs:appendix}):
%%%%%%%%%%%%%%%%%%%%%%%% EQUATION %%%%%%%%%%%%%%%%%%%%%%%%
\begin{widetext}
    \begin{equation}
        \begin{aligned}
            g = &\frac{1}{\beta} \ln \Lambda -\frac{1}{\beta}
            [(1-x)\ln(1-x) + x\ln(x) -x_{AA}\ln(x_{AA})-x_{BB}\ln(x_{BB}) \\
            &-x_{AB}\ln(x_{AB}/2) + x_{AA}\ln(\varphi_{AA}) + x_{BB}\ln(\varphi_{BB}) + x_{AB}\ln(\varphi_{AB})],
            \label{eq:g}
        \end{aligned}
    \end{equation}
\end{widetext}
%%%%%%%%%%%%%%%%%%%%%%%% EQUATION %%%%%%%%%%%%%%%%%%%%%%%%
where
%%%%%%%%%%%%%%%%%%%%%%%% EQUATION %%%%%%%%%%%%%%%%%%%%%%%%
\begin{equation}
\begin{aligned}
    x_\text{A} &\equiv \frac{N_A}{N}, \\
    x_\text{B} &\equiv \frac{N_B}{N}, \\
    x &\equiv x_B = 1-x_A, \\
    x_{AA} &= \frac{(x_A-x_B) - 2x_A\gamma + \sqrt{(x_A-x_B)^2 + 4x_Ax_B\gamma}}{2\left(1 - \gamma\right)},\\
    x_{BB} &= \frac{-(x_A-x_B) - 2x_B\gamma + \sqrt{(x_A-x_B)^2 + 4x_Ax_B\gamma}}{2\left(1 - \gamma\right)},\\
    x_{AB} &= 2(x_A-x_{AA}), \\
    \gamma &\equiv \frac{\varphi_{AA}\varphi_{BB}}{\varphi_{AB}^2}.
\end{aligned}
\end{equation}
%%%%%%%%%%%%%%%%%%%%%%%% EQUATION %%%%%%%%%%%%%%%%%%%%%%%%
The $x_{ij}$ quantities represent a neighbor fraction: a small $x_{AA}$ value means that few $A$ particles have an $A$ neighbor. Their derivation is given in Appendix~\ref{sec:gibbs:appendix}.

%%%%%%%%%%%%%%%%%%%%%%%%%%%%%% 1D - RESULTS%%%%%%%%%%%%%%%%%%%%%%%%%%
\subsection{Results}
%%%%%%%%%%%%%%%%%%%%%%%%%%%%%% 1D - RESULTS%%%%%%%%%%%%%%%%%%%%%%%%%%
In order to test our assumption that a solute-solvent attraction leads to an increase in the TMD, we introduce the following pair potentials:
%%%%%%%%%%%%%%%%%%%%%%%% EQUATION %%%%%%%%%%%%%%%%%%%%%%%%
\begin{align}
    V_{AA} (r) &= \begin{cases}
        0, & \text{if } r > d_{AA}; \\
        a'_{AA}, & \text{if } c_{AA}<r<d_{AA}; \\
        -a_{AA}, & \text{if } b_{AA}<r<c_{AA}; \\
        \infty, & \mbox{if } r<b_{AA}. \end{cases} \\
    V_{iB} (r) &= \begin{cases}
        0, & \text{if } r >c_{iB}; \\
        -a_{iB}, & \text{if } b_{iB}<r<c_{iB}; \\
        \infty, & \mbox{if } r<b_{iB} \;. \end{cases}
    \label{eq:hard_cs_potential}
\end{align}
%%%%%%%%%%%%%%%%%%%%%%%% EQUATION %%%%%%%%%%%%%%%%%%%%%%%%
The chosen solvent-solvent ($AA$) interaction is a CS potential followed by an attractive well. This type of two length-scales potential guarantees the presence of the temperature of maximum density for the pure solvent system~\cite{hemmer1972coresoftened,gribova2009,barbosa2013,wilding2002phase,almarza2009phase,camp2003structure,jagla1998phase}.

The solvent-solute ($AB$) interaction is a van der Waals attraction well. This potential captures the potential hydrogen bond in the solution of water and alcohol. The size of the solute is larger than the solvent as in the case of the alcohols in water.
%%%%%%%%%%%%%%%%%%%%%%%% FIGURE 1 %%%%%%%%%%%%%%%%%%%%%%%%
\begin{figure}[t]
    \centering
    \includegraphics{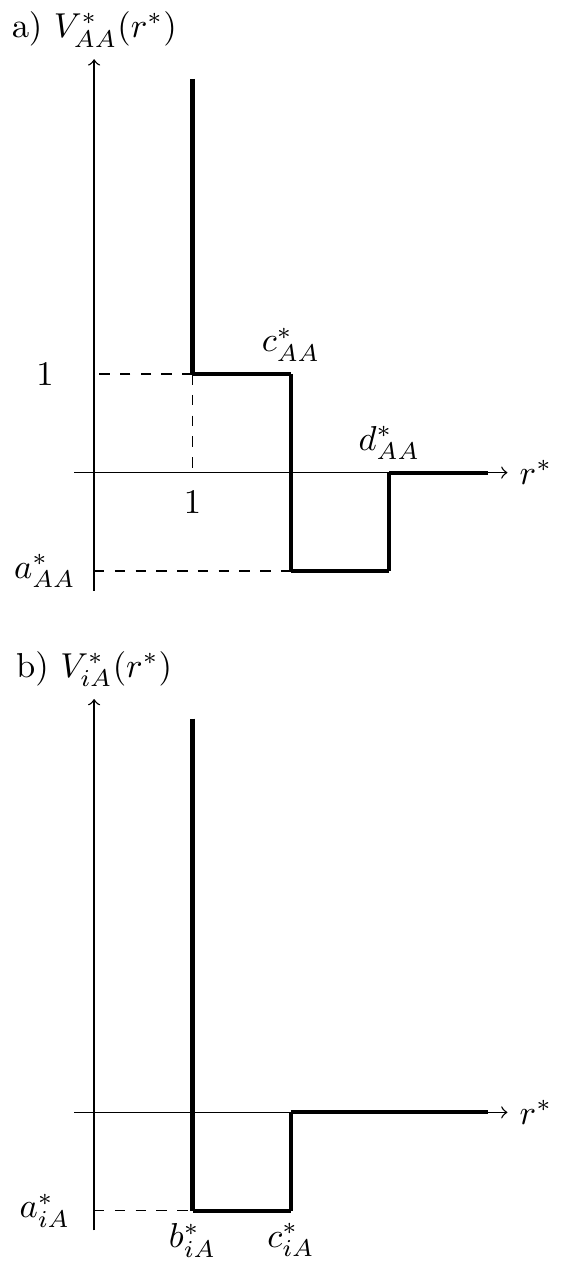}
    \caption{(a) solvent-solvent, (b) solvent-solute/solute-solute interaction reduced potentials as a function of reduced distance.}
    \label{fig:1d_potential}
\end{figure}
%%%%%%%%%%%%%%%%%%%%%%%% FIGURE 1 %%%%%%%%%%%%%%%%%%%%%%%%
Finally, the solute-solute ($BB$) interaction is simply a repulsive core. Here we are not including the potential hydrogen bond between two alcohols since we assume it is not related to the mechanism for the increase of the TMD. We performed the same simulations with a weakly attractive solute-solute interaction and verified no difference in the TMD change from the results presented here.

We consider the following values for the reduced parameters, expressed in terms of the energy $a'_{AA}$ and length $b_{AA}$:
%%%%%%%%%%%%%%%%%%%%%%%% EQUATION %%%%%%%%%%%%%%%%%%%%%%%%
\begin{equation*}
\begin{aligned}
    a^{*}_{AA} & = \frac{a_{AA}}{a'_{AA}} = 0.5, c^{*}_{AA} = \frac{c_{AA}}{b_{AA}} =1.7, d^{*}_{AA} = \frac{d_{AA}}{b_{AA}} = 1.8; \\
    b^{*}_{AB} &= \frac{b_{AB}}{b_{AA}} = 2.3, c^{*}_{AB} = \frac{c_{AB}}{b_{AA}} = 2.4; \\
    a^{*}_{BB} & = \frac{a_{BB}}{a'_{AA}} = 0, b^{*}_{BB} = c^{*}_{BB} = \frac{c_{BB}}{b_{AA}} = 2.4 \;.
\end{aligned}
\end{equation*}
%%%%%%%%%%%%%%%%%%%%%%%% EQUATION %%%%%%%%%%%%%%%%%%%%%%%%
These parameters were found through numerical testing to exhibit a TMD for the pure system~\cite{rizzatti2018oned} and for the mixture. The parameter $a^{*}_{AB} = \frac{a_{AB}}{a'_{AA}}$ is free, to identify different solute types: structure maker or structure breaker. The parameters for the solute-solute interaction reflects the purely repulsive potential with $a^{*}_{BB} = 0$ and $c^{*}_{BB} = b^{*}_{BB}$. Figure~\ref{fig:1d_potential} illustrates the interaction potentials $AA$ and $iA$, where $i \in \{A,B\}$, in reduced units.

\begin{figure}[t]
   \centering
  \includegraphics{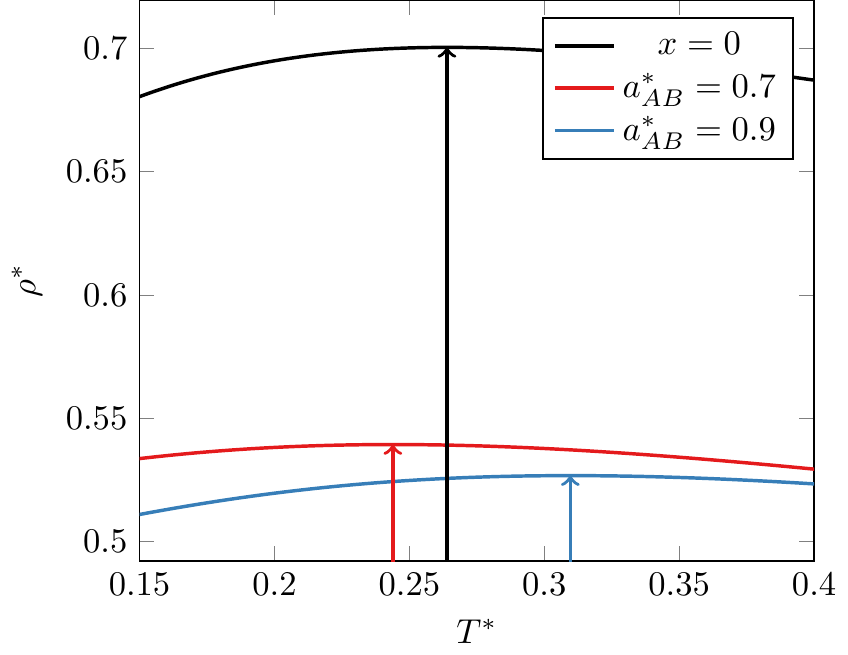}
    \caption{Density as a function of temperature at $P^{*}=2$ for: pure solvent (black line), $a^{*}_{AB}=0.7$ (red line) and $a^{*}_{AB}=0.9$ (blue line) at $x=0.30$. Vertical arrows show how the increase in density changes the TMD.}
     \label{fig:1d_tmd}
 \end{figure}

From Equation~\ref{eq:g} we obtain the free energy and the density profile for different pressures, temperatures and solute fractions. The results were also calculated in terms of reduced units of energy, $a'_{AA}$, and length, $b_{AA}$:
%%%%%%%%%%%%%%%%%%%%%%%% EQUATION %%%%%%%%%%%%%%%%%%%%%%%%
\begin{equation}
\begin{aligned}
    P^{*} &= P \frac{b_{AA}}{a'_{AA}} , \\
    T^{*} &= \frac{k_B T}{a'_{AA}} , \\
    \rho^{*} &= \rho \; b_{AA}\;.
\end{aligned}
\end{equation}
%%%%%%%%%%%%%%%%%%%%%%%% EQUATION %%%%%%%%%%%%%%%%%%%%%%%%
Figure~\ref{fig:1d_tmd} shows the density versus temperature diagram at fixed pressure of the one dimensional system with particles interacting through the potential shown in Figure~\ref{fig:1d_potential}. For the pure system, $x=0$, the density at low temperatures increases with the increase of the temperature. This behavior resembles the observed density of water~\cite{kell1975}.
%%%%%%%%%%%%%%%%%%%%%%%%  FIGURE 2 %%%%%%%%%%%%%%%%%%%%%%%%
%%%%%%%%%%%%%%%%%%%%%%%% FIGURE 2 %%%%%%%%%%%%%%%%%%%%%%%%

In the case of pure water, the TMD is understood as follows. At low temperatures, the water molecules form a hydrogen bond network with few non bonded molecules, creating an ``expanded structure''. As the temperature is increased at constant pressure, some hydrogen bonds are broken and the number of non bonded molecules increases. To keep the pressure fixed these non bonded molecules are closer than the bonded molecules, forming a ``compacted structure'', increasing the density. At a certain threshold temperature, the number of non bonded molecules is large enough to allow for some of them to be far apart, decreasing the density with the increase of the temperature. This threshold defines the TMD.

The idea of liquid water being a mixture of two states, expanded and compacted structures, is not new~\cite{davis1965}. But the link between them and all the anomalous behavior of water which led to the hypothesis of two liquid phases for water~\cite{poole1992} is still being explored, both in phenomenological~\cite{shi2020} and more general models~\cite{ciach2008simple,anisimov2018thermodynamics,cerdeirina2019water,caupin2021minimal}.

Specifically in our one dimensional pure solvent system, as in other core softened potentials~\cite{franzese2007,alan2008,barbosa2013}, the density anomaly emerges not from bonded and non bonded molecules as in water but from the competition between particles arranging themselves in two length scales: the attractive well, $c_{AA}<r<d_{AA}$, and the repulsive shoulder-like, $b_{AA}<r<c_{AA}$, as shown in Figure~\ref{fig:1d_potential}(a). At low temperatures the majority of $A$ particles are located at the attractive well length scale, forming the ``expanded structure''. As the temperature is increased at fixed pressure, some particles are at shoulder length scale, forming the ``compacted structure''~\cite{rizzatti2018oned,barbosa2013}. For temperatures above a certain threshold, particles can be at distances beyond the attractive well length scale, $r>d_{AA}$, decreasing the density. This two length scales mechanism leads to the temperature of maximum density as illustrated by the $x=0$ case in Figure~\ref{fig:1d_tmd}~\cite{barbosa2013}.

There is a TMD for a range of pressure values. As pressure is increased, the expanded structures exhibits a lower percentage of particles at the attractive length scale what move the threshold temperature to lower values. Therefore, the increase of pressure reduces the TMD. In parallel, as the attractive potential becomes deeper, the percentage of particles at the attractive scale increases and the TMD increases~\cite{dasilva2010}.

A non-interacting or weakly interacting solutes disturbs the expanded structure of the solvent, decreasing the number of solvent particles at the attractive length scale. Consequently, the temperature of maximum density is lower when compared with the pure solvent system as shown by the $a^{*}_{AB}=0.7$ case in Figure~\ref{fig:1d_tmd}(a). Similar behavior was observed for a pure hard sphere solute in a lattice model~\cite{szortyka2012}.

In the case of systems with strong attractive solute-solvent interaction, the solute molecules, instead of disrupting the hydrogen bonds, tend to occupy locations between two  solvent molecules, the interstitial vacancies~\cite{subramanian2011}. Hence, the TMD increases when compared with the pure system as illustrated by the $a^{*}_{AB}=0.9$ case in Figure~\ref{fig:1d_tmd}.

In order to determine the behavior of the TMD for different solute-solvent interaction potentials we calculated the $\Delta T^{*}_{\text{MD}}(x)$, namely
%%%%%%%%%%%%%%%%%%%%%%%% EQUATION %%%%%%%%%%%%%%%%%%%%%%%%
\begin{equation}
    \Delta  T^{*}_{\text{MD}}(x) =  T^{*}_{\text{MD}}(x) -  T^{*}_{\text{MD}}(0) \ ,
\end{equation}
%%%%%%%%%%%%%%%%%%%%%%%% EQUATION %%%%%%%%%%%%%%%%%%%%%%%%
which is the difference between the temperature of maximum density of the system with a certain concentration $x$ of solute and the TMD of the pure $A$ system. A positive $\Delta T^{*}_{\text{MD}}(x)$ implies that the solute is a structure maker, while a negative $\Delta T^{*}_{\text{MD}}(x)$ means that the solute is a structure breaker.

\begin{figure}
    \centering
    \includegraphics{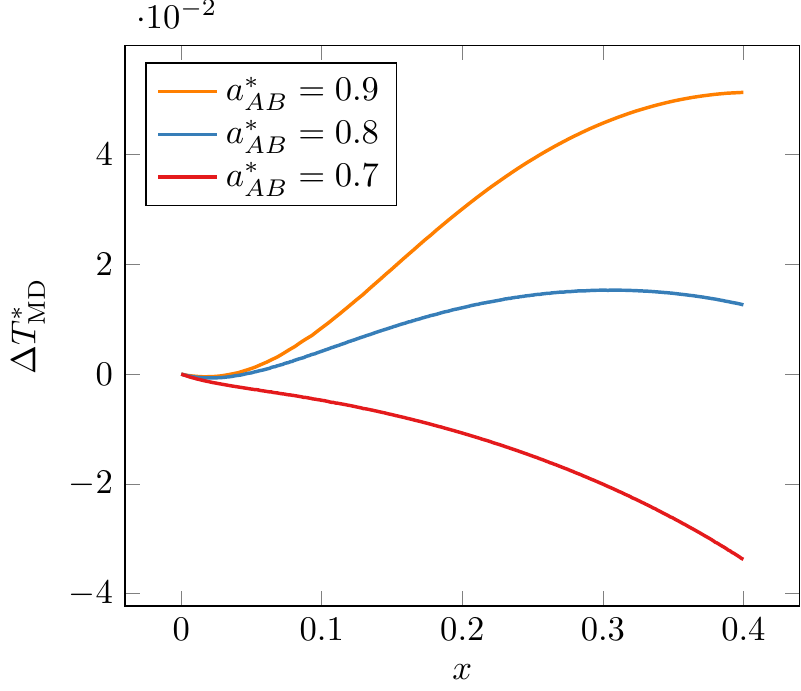}
    \caption{Change in the temperature of maximum density $\Delta T^{*}_{\text{MD}}$ as a function of solute concentration at $P^{*}=2$ for $a^{*}_{AB}=0.9$ (orange), 0.8 (blue), 0.7 (red).}
    \label{fig:1d_dTMD_change_aAB}
\end{figure}

Figure~\ref{fig:1d_dTMD_change_aAB} shows the behavior of $\Delta T^{*}_{\text{MD}}$ as a function of solute concentration, $x$, for $a^{*}_{AB}=0.7$, 0.8 and 0.9 at a pressure of $P^{*}=2$. The graph shows that the enhancement of the solute-solvent attraction leads to an increase of the $\Delta T^{*}_{\text{MD}}(x)$ for $a^{*}_{AB} > 0.756$. This increase in the $\Delta T^{*}_{\text{MD}}(x)$ is consistent with observations for lattice models in the case of large solute-solvent attraction~\cite{girardi2015} and with experimental results for structure-maker solutes where $\Delta T^{*}_{\text{MD}}(x)>0$~\cite{wada1962}.
%%%%%%%%%%%%%%%%%%%%%%%% FIGURE 3 %%%%%%%%%%%%%%%%%%%%%%%%
%%%%%%%%%%%%%%%%%%%%%%%% FIGURE 3  %%%%%%%%%%%%%%%%%%%%%%%%
 
In order to understand the TMD increase for large solute-solvent attraction we looked at how the solvent and solute are organized. If the solvent $A$ and the solute $B$ particles do not correlate but exhibit a random distribution, the probabilities of having two solvent-solvent $AA$ or solvent-solute $AB$ as first neighbors are given respectively by
%%%%%%%%%%%%%%%%%%%%%%%% EQUATION %%%%%%%%%%%%%%%%%%%%%%%%
\begin{equation}
    \begin{aligned}
        x_{AA}^{\text{random}}&= \frac{N_A^2}{N^2} = (1-x)^2, \\
        x_{AB}^{\text{random}}&= 2\frac{N_A N_B}{N^2} = 2x (1-x) \;.
        \label{eq:random}
    \end{aligned}
\end{equation}
%%%%%%%%%%%%%%%%%%%%%%%% EQUATION %%%%%%%%%%%%%%%%%%%%%%%%

For the non-random system these probabilities are given by the interaction fraction $x_{AA}$ and $x_{AB}$: if we randomly pick a particle of type $i$, the probability of a particle $j$ being next to it is $N_{ij}/N \equiv x_{ij}$. Therefore, the impact of the fluctuation effects can be measured by the ratio between the non-random values, $x_{AA}$ and $x_{AB}$, with the random approximations, $x_{AA}^{\text{random}}$ and $x_{AB}^{\text{random}}$, determined by Equation~\ref{eq:random}, leading to
%%%%%%%%%%%%%%%%%%%%%%%% EQUATION %%%%%%%%%%%%%%%%%%%%%%%%
\begin{equation}
    \begin{aligned}
        x_{AA}^{\text{corr}} &= \frac{x_{AA}}{x_{AA}^{\text{random}}} =\frac{x_{AA}}{(1-x)^2}, \\
        x_{AB}^{\text{corr}} &= \frac{x_{AB}}{x_{AB}^{\text{random}}} =\frac{x_{AB}}{2x(1-x)}\;.
        \label{eq:corr}
    \end{aligned}
\end{equation}
%%%%%%%%%%%%%%%%%%%%%%%% EQUATION %%%%%%%%%%%%%%%%%%%%%%%%

\begin{figure}[t]
    \centering
    \includegraphics[width=8.5cm]{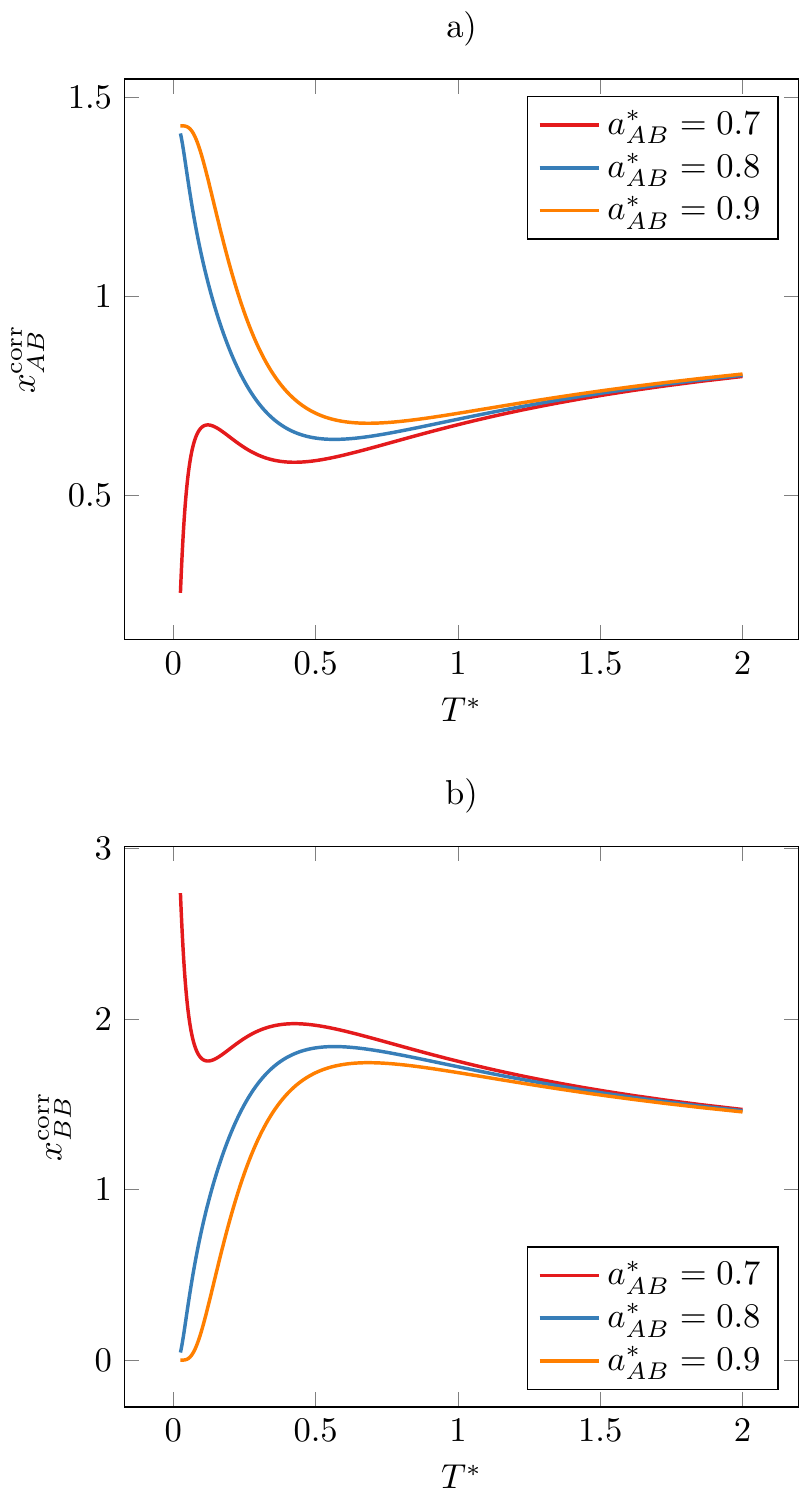}
    \caption{(a) Interaction correlations for different $AB$ well depths as a function of temperature. The plots were made with a concentration $x=0.30$ and $P^{*}=2$ for $a^{*}_{AB}=0.9$ (orange), 0.8 (blue), 0.7 (red).}
    \label{fig:1d_change_aAB_density}
\end{figure}

For very high temperatures we would expect that $x_{AA}^{\text{corr}}=1$, $x_{AB}^{\text{corr}}=1$ and $x_{BB}^{\text{corr}}=1$. Figure~\ref{fig:1d_change_aAB_density}(a) shows  $x^{\text{corr}}_{AB}$ versus temperature for the $a^{*}_{AB}=0.7$, 0.8 and 0.9 cases at fixed pressure $P^{*}=2.0$ and solvent concentration $x=0.3$. At low temperatures and high solute-solvent attraction, $a^{*}_{AB}=0.8, 0.9$,  $x^{\text{corr}}_{AB}$ shows that the solute is bounded to the solvent while for $a^{*}_{AB}=0.7$ fewer solute molecules are bounded to the solvent. 

Figure~\ref{fig:1d_change_aAB_density}(b) indicates that for high solute-solvent attraction few solute molecules form pairs, while for low attraction a larger number of them form pairs. Therefore, the unpaired solute molecules are caged by the solvent. This caged solute forms a structure which needs more temperature to be disrupted, which increases the TMD. This mechanism of trapping the solute in a solvent shell was observed in atomistic models for water, where it was identified the solute is confined in clathrates~\cite{lomba2022-experiment}.

%%%%%%%%%%%%%%%%%%%%%%%% FIGURE 4 %%%%%%%%%%%%%%%%%%%%%%%%
%%%%%%%%%%%%%%%%%%%%%%%% FIGURE 4 %%%%%%%%%%%%%%%%%%%%%%%%

\begin{figure}[t]
    \centering
    \includegraphics[width=8.5cm]{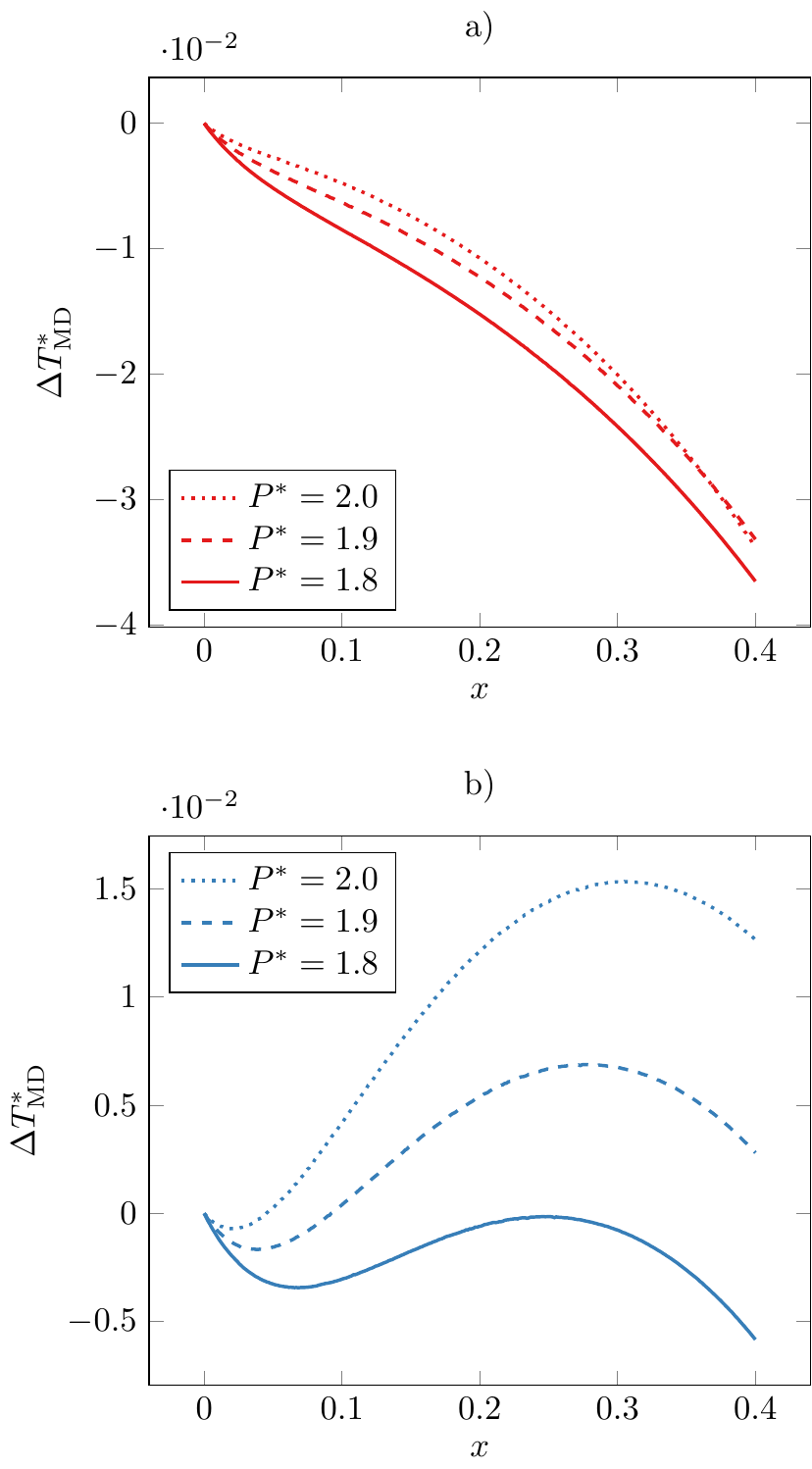}
    \caption{Change in the temperature of maximum density $\Delta T^{*}_{\text{MD}}$ as a function of solute concentration at $P^{*}=1.8$ (solid line), $P^{*}=1.9$ (dashed line) and $P^{*}=2.0$ (dotted line) for (a) $a^{*}_{AB}=0.7$ and (b) $a^{*}_{AB}=0.8$.}
    \label{fig:1d_dTMD_change_pressure_aAB}
\end{figure}

Next, we compute how the behavior of $\Delta T^{*}_{\text{MD}}(x)$ is impacted by the pressure for both structure breakers and structure makers solutes. In the pure system and in the presence of the solute, the TMD decreases with increasing pressure. However,  Figure~\ref{fig:1d_dTMD_change_pressure_aAB}(a) shows increasing pressure increases $\Delta T^{*}_{\text{MD}}(x)$ for both structure breakers and structure makers.

However, only for structure makers $\Delta T^{*}_{\text{MD}}(x)$ becomes positive. This indicates that pressure helps to disrupt the ``expanded structure'' in both the pure system and in the presence of solute. For $a^{*}_{AB}=0.8, 0.9$ the solute is caged by the solvent and disrupting this structure requires larger temperatures, increasing the TMD. This behavior is consistent with what was observed in recent experiments and atomistic simulations~\cite{lomba2022-experiment}: a lower pressure shifts the $\Delta T^{*}_{\text{MD}}(x)$ curve downwards.
%%%%%%%%%%%%%%%%%%%%%%%% FIGURE 5 %%%%%%%%%%%%%%%%%%%%%%%%
%%%%%%%%%%%%%%%%%%%%%%%% FIGURE 5 %%%%%%%%%%%%%%%%%%%%%%%%
%%%%%%%%%%%%%%%%%%%%%%%% 3D %%%%%%%%%%%%%%%%%%%%%%%%
\section{Three Dimensional System}
\label{sec:3d}
%%%%%%%%%%%%%%%%%%%%%%%% 3D %%%%%%%%%%%%%%%%%%%%%%%%

%%%%%%%%%%%%%%%%%%%%%%%% 3D - METHODS %%%%%%%%%%%%%%%%%%%%%%%%
\subsection{Model and Methods}
%%%%%%%%%%%%%%%%%%%%%%%% 3D - METHODS %%%%%%%%%%%%%%%%%%%%%%%%
We performed MD simulations with LAMMPS~\cite{LAMMPS} in the NPT ensemble with the Nose-Hoover thermostat and barostat. The simulated system consists of $N = 1000$ particles, with $N_B=xN$ being the number of solute particles. The two types of particles were placed in a three dimensional lattice with periodic boundary conditions and $10^6$ time steps were run for the system to reach equilibrium. We then used another $2\times 10^6$ steps for averaging the thermodynamic quantities, with a $\delta t^{*} = 0.005$ time step.

We created a continuous potential which resembles the length scales of the one dimensional case as described below. The physical quantities are calculated by Molecular Dynamics simulation. All results are displayed in reduced units, where
%%%%%%%%%%%%%%%%%%%%%%%% EQUATION %%%%%%%%%%%%%%%%%%%%%%%%
\begin{align*}
    \epsilon^{*}_{ij} &= \frac{\epsilon_{ij}}{\epsilon_{AA}}, \\
    V^{*}_{ij} &= \frac{V_{ij}}{\epsilon_{AA}}, \\
    P^{*} &= P \frac{\sigma^3_{AA}}{\epsilon_{AA}}, \\
    T^{*} &= \frac{k_B T}{\epsilon_{AA}}, \\
    \rho^{*} &= \rho \; \sigma^3_{AA} \; .
\end{align*}
%%%%%%%%%%%%%%%%%%%%%%%% EQUATION %%%%%%%%%%%%%%%%%%%%%%%%

In the three dimensional system the solvent-solvent ($AA$) interaction is given by a two length scale potential, namely:
%%%%%%%%%%%%%%%%%%%%%%%% EQUATION %%%%%%%%%%%%%%%%%%%%%%%%
\begin{widetext}
    \begin{align*}
        V_{AA} (r) &= 4 \epsilon_{AA} \left[\left(\frac{\sigma_{AA}}{r}\right)^{12}-\left(\frac{\sigma_{AA}}{r}\right)^{6}\right] +\sum_{l=0}^{1} u_l \epsilon_{AA} \exp \left[-\frac{1}{c_{l}^{2}}\left(\frac{r-r_{l}}{\sigma_{AA}}\right)^{2}\right],
    \end{align*}
\end{widetext}
%%%%%%%%%%%%%%%%%%%%%%%% EQUATION %%%%%%%%%%%%%%%%%%%%%%%%
where we used $u_0 = 5$, $u_1 = -0.75$, $c_0 = 1$, $c_1 = 0.5$, $r^{*}_0 = 0.7$ and $r^{*}_1 = 2.5$. This potential for the solvent interaction  presents two length scales: one attractive at $r^*\approx 2.5$ and one repulsive shoulder-like at $r^* \approx 1.2$. The pure system exhibits a TMD~\cite{alan2006,dasilva2010}.

\begin{figure}[t]
    \centering
    \includegraphics{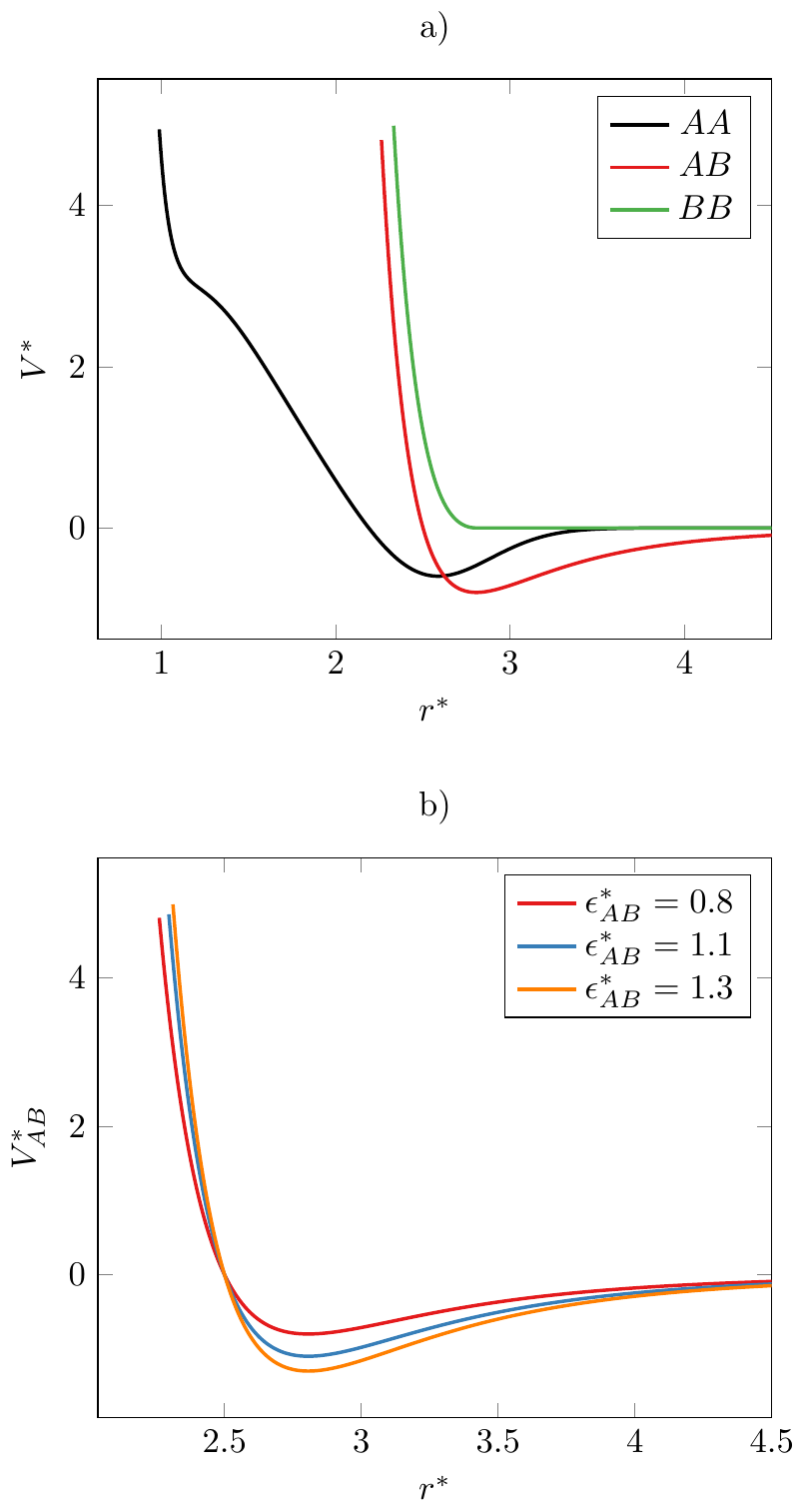}
    \caption{Interaction potential versus distance for the (a) solvent-solvent (black line), solute-solvent  (red line) and solute-solute (green line) interactions (b) solvent-solute for $\epsilon^{*}_{AB}=0.8$ (red line), $\epsilon^{*}_{AB}=1.1$ (blue line) and $\epsilon^{*}_{AB}=1.3$ (yellow line).}
    \label{fig:potential_3d}
\end{figure}

The solute-solvent ($AB$) interaction is a Lennard-Jones well with $\sigma^{*}_{AB}=2.5$, with $\epsilon^{*}_{AB}$ kept a free parameter.  Finally, the solute-solute ($BB$) interaction is a purely repulsive Weeks-Chandler-Andersen~\cite{wca} potential with $\epsilon^{*}_{BB}=1.2$ and $\sigma^{*}_{BB} = 2.5$. These potentials, in units of $\epsilon_{AA}$, versus distance, in units of $\sigma_{AA}$, are represented in Figure~\ref{fig:potential_3d}(a). Here we used a cut-off radius $r^{*}_c=5$ and the figure is plotted for $\epsilon^{*}_{AB}=0.8$. Note that we do not adopt the Lorentz-Berthelot approximation, which seems to be relevant to the increase of $\Delta T^{*}_{\text{MD}}(x)$.
%%%%%%%%%%%%%%%%%%%%%%%% FIGURE 6 %%%%%%%%%%%%%%%%%%%%%%%%
%%%%%%%%%%%%%%%%%%%%%%%% FIGURE 6 %%%%%%%%%%%%%%%%%%%%%%%%

%%%%%%%%%%%%%%%%%%%%%%%% 3D RESULTS %%%%%%%%%%%%%%%%%%%%%%%%
\subsection{Results}
%%%%%%%%%%%%%%%%%%%%%%%% 3D RESULTS %%%%%%%

The 1D case suggests that $\Delta T^{*}_{\text{MD}}(x)>0$ occurs if the solute-solvent interaction becomes attractive enough. Now, we check if this is the case in three dimensions by analyzing the change in TMD as a function of solute concentration and interaction strength in the 3D model. Three different $AB$ attractions, $\epsilon^{*}_{AB} = 0.8$, 1.1 and 1.3, from less to more attractive, respectively, are shown in Figure~\ref{fig:potential_3d}(b).

\begin{figure}[t]
    \centering
    \includegraphics{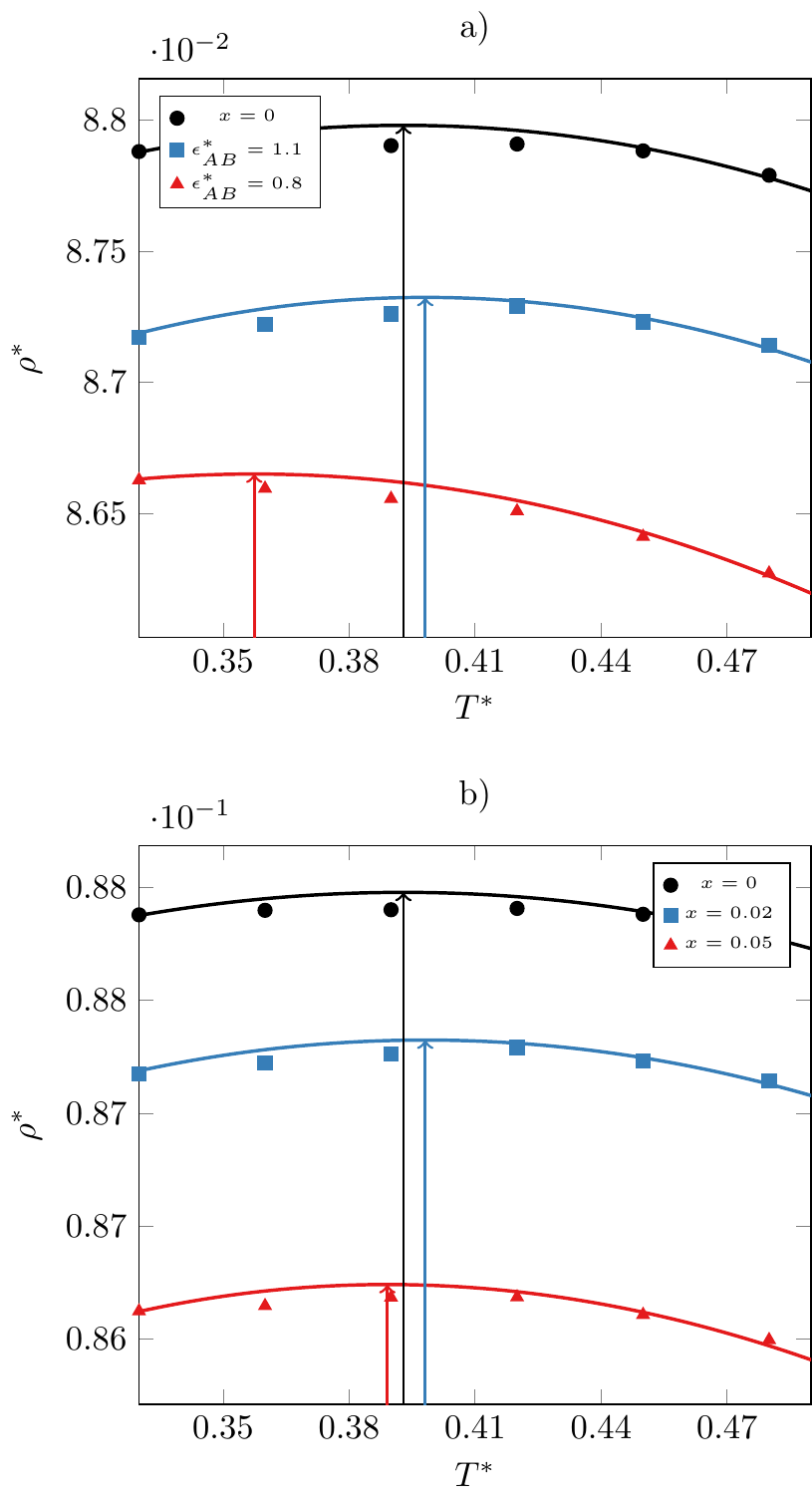}
    \caption{Density as a function of temperature at $P^{*}=0.13$ for (a) pure solvent (black circles), $\epsilon^{*}_{AB}=1.1$ (blue squares) and $\epsilon^{*}_{AB}=0.8$ (red triangles), both at $x=0.02$; (b) pure solvent (black circles), $\epsilon^{*}_{AB}=1.1$ with $x=0.02$ (blue squares) and $\epsilon^{*}_{AB}=1.1$ with $x=0.05$ (red triangles). Vertical arrows indicate the TMD. Markers indicate some of the simulation points. Curves are third degree polynomial fits.}
    \label{fig:3d_density_comparison}
\end{figure}

Figure~\ref{fig:3d_density_comparison}(a) illustrates the density versus temperature at $P^{*}=0.13$ for $\epsilon^{*}_{AB}=0.8$ and $1.1$ with $x=0.02$. The pure case ($x=0$) is also shown. The temperature of maximum density of the pure solvent system is larger than the TMD for the $\epsilon^{*}_{AB}= 0.8$ case and lower than the TMD for the more attractive, $\epsilon^{*}_{AB}=1.1$, system.

Figure~\ref{fig:3d_density_comparison}(b) shows the density versus temperature for $\epsilon^{*}_{AB}=1.1$ at different solute concentrations, $x=0.02$ and $x=0.05$, compared with the pure solvent system, $x=0$. The TMD of the $x=0.02$ system is larger than the TMD of the pure system, while the TMD of the pure solvent system is higher than the TMD of the $x=0.05$ case. That is, $\Delta T^{*}_{\text{MD}}(x=0.05) < \Delta T^{*}_{\text{MD}}(x=0) < \Delta T^{*}_{\text{MD}}(x=0.02)$. Analyzing different concentrations we observe that if the solvent-solute attraction is strong enough, $\epsilon^{*}_{AB}=1.1$ for instance, the addition of a small amount of the solute leads to an increasing TMD. However, the addition of a larger quantity of solute, such as $x=0.05$, makes the TMD decrease. For a less attractive solute-solvent interaction, $\epsilon^{*}_{AB}=0.8$ for instance, the addition of the solute only decreases the TMD when compared with the pure solvent system.
%%%%%%%%%%%%%%%%%%%%%%%% FIGURE 7 %%%%%%%%%%%%%%%%%%%%%%%%
%%%%%%%%%%%%%%%%%%%%%%%% FIGURE 7 %%%%%%%%%%%%%%%%%%%%%%%%

\begin{figure}[t]
    \centering
    \includegraphics{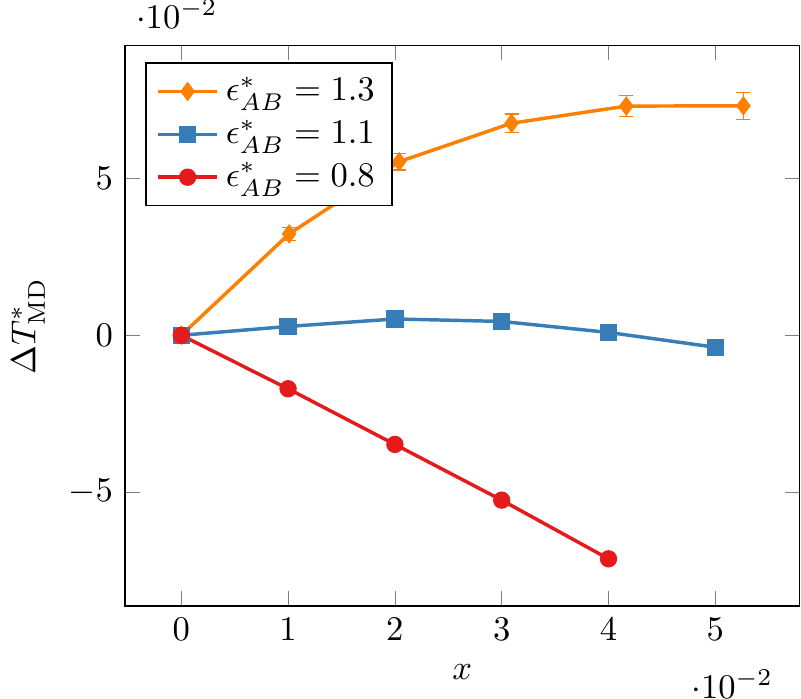}
    \caption{Change in the temperature of maximum density as a function of solute concentration at $P^{*}=0.13$ for different values of $\epsilon^{*}_{AB}$: $\epsilon^{*}_{AB}=0.8$ (red line with circles), $\epsilon^{*}_{AB}=1.1$ (blue line with squares) and $\epsilon^{*}_{AB}=1.3$ (yellow line with diamonds).}
    \label{fig:3d_change_v0AB_dtmd}
\end{figure}

In order to understand how the TMD changes with solute fraction, we plot $\Delta T^{*}_{\text{MD}}(x)$ in Figure~\ref{fig:3d_change_v0AB_dtmd}. We observe that for more attractive systems, $\epsilon^{*}_{AB}=1.1$ and $1.3$,  $\Delta T^{*}_{\text{MD}}(x)$ increases with the increase of $x$ at low solute concentrations, while for a lower attraction between solute and solvent, $\epsilon^{*}_{AB}=0.8$, $\Delta T^{*}_{\text{MD}}(x)$ is negative for all values of $x$. This result is consistent with observations for alcohols~\cite{wada1962}. In particular, it indicates that the more attractive solute generates a larger $\Delta T^{*}_{\text{MD}}(x)$ for the same fraction of solute.
%%%%%%%%%%%%%%%%%%%%%%%% FIGURE 8 %%%%%%%%%%%%%%%%%%%%%%%%
%%%%%%%%%%%%%%%%%%%%%%%% FIGURE 8 %%%%%%%%%%%%%%%%%%%%%%%%

Figure~\ref{fig:3d_change_v0AB_dtmd} suggests that a major factor for structure makers is the attractive interaction with the solute. In order to understand if the same mechanism of the solute being caged by solvent, as observed in the one dimensional case, also appears in this model, we computed the radial distribution function.

\begin{figure}[t]
    \centering
    \includegraphics{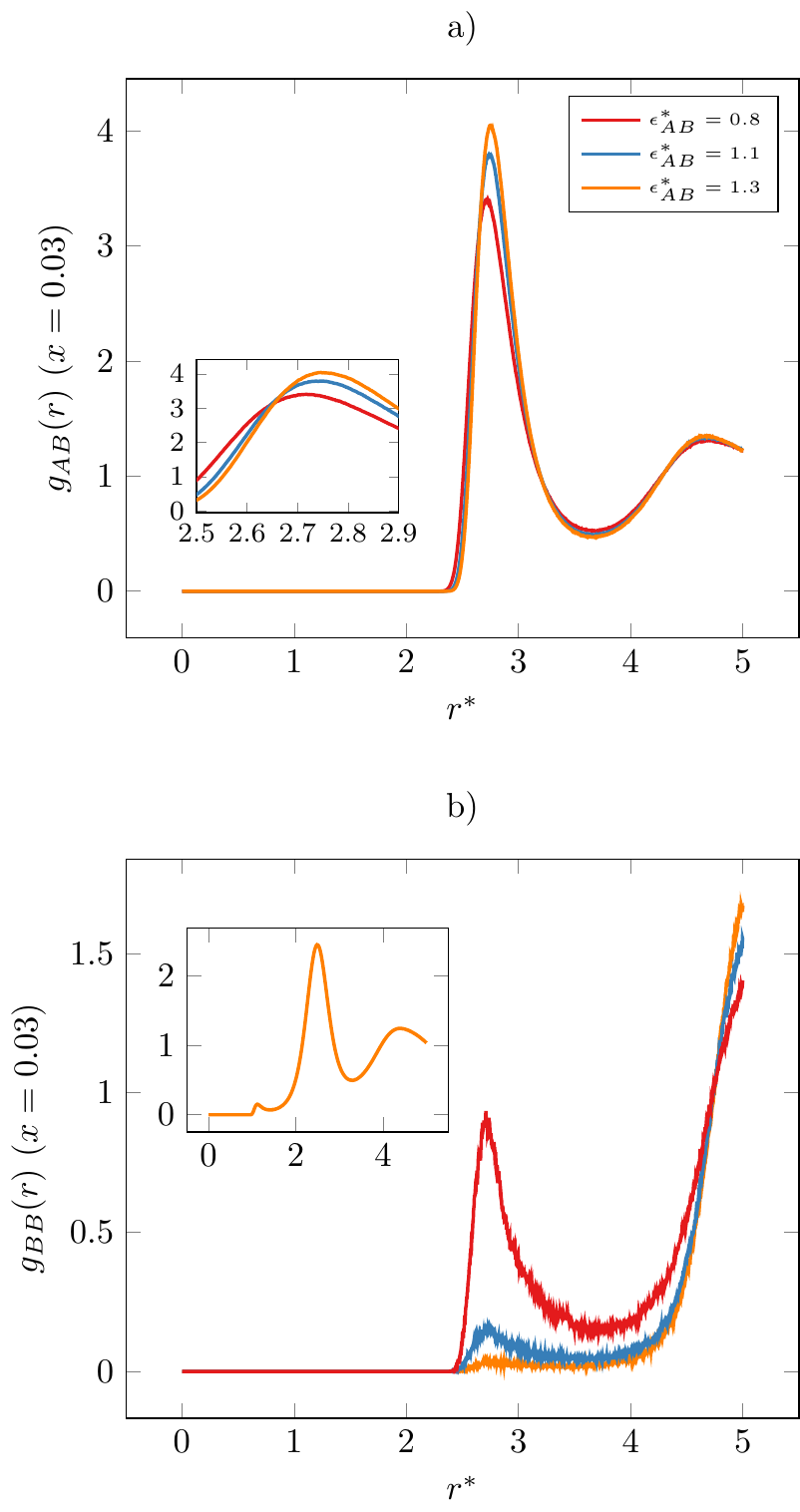}
    \caption{Radial distribution function for $\epsilon^{*}_{AB}=0.8$ (red line), $\epsilon^{*}_{AB}=1.1$ (blue line)$ and \epsilon^{*}_{AB}=1.3$ (yellow line), at $P^{*}=0.13$, $T^{*}=0.40$ and $x=0.03$ for (a) solvent-solute, $g_{AB}$, and (b) solute-solute, $g_{BB}$. The inset in (a) zooms in the $r^{*} \approx 2.7$ region. The inset in (b) is the solvent-solvent RDF at $P^{*}=0.13$, $T^{*}=0.40$ and $x=0.03$.}
    \label{fig:3d_change_v0AB_rdf}
\end{figure}

Figure~\ref{fig:3d_change_v0AB_rdf}(a) shows the solute-solvent $g_{AB} (rr^{*})$ for different levels of solute-solvent attraction, $\epsilon^{*}_{AB}=0.8$, 1.1 and 1.3. The distribution of particles indicates that, as the solute-solvent interaction becomes more attractive, the first coordination shell between solute and solvent becomes more populated. In addition, the graph also shows that the second coordination shell of solute-solvent molecules is distant from the first coordination shell by the attractive length scale, $r^* \approx 2.5$. This suggests that each solute is surrounded by the solvent's ``expanded structure''.

Figure~\ref{fig:3d_change_v0AB_rdf}(b) shows the solute-solute $g_{BB} (r^{*})$ for different levels of solute-solvent attraction, $\epsilon^{*}_{AB}=0.8$, 1.1 and 1.3. The distribution indicates that, as the interaction between solute and solvent becomes more attractive, less solute pairs are observed. This result is consistent with the observations of our one dimensional model and with the idea that the solute molecules at low concentrations is caged by bounded water. In this same figure, the inset shows $g_{AA} (r^{*})$ for $\epsilon^{*}_{AB}=1.3$. The solvent-solvent RDF shows a two-peak structure at $r^{*} \approx 1.3 $ and $r^{*} \approx 2.5$, which is a signature of water's anomalous behaviors~\cite{alan2008,furlan2017}. A decrease in $\epsilon^{*}_{AB}$ lowers the first peak and increases the second.

%%%%%%%%%%%%%%%%%%%%%%%% FIGURE 9 %%%%%%%%%%%%%%%%%%%%%%%%
%%%%%%%%%%%%%%%%%%%%%%%% FIGURE 9 %%%%%%%%%%%%%%%%%%%%%%%%

%%%%%%%%%%%%%%%%%%%%%%%% FIGURE 10 %%%%%%%%%%%%%%%%%%%%%%%%
\begin{figure}[t]
    \centering
    \includegraphics{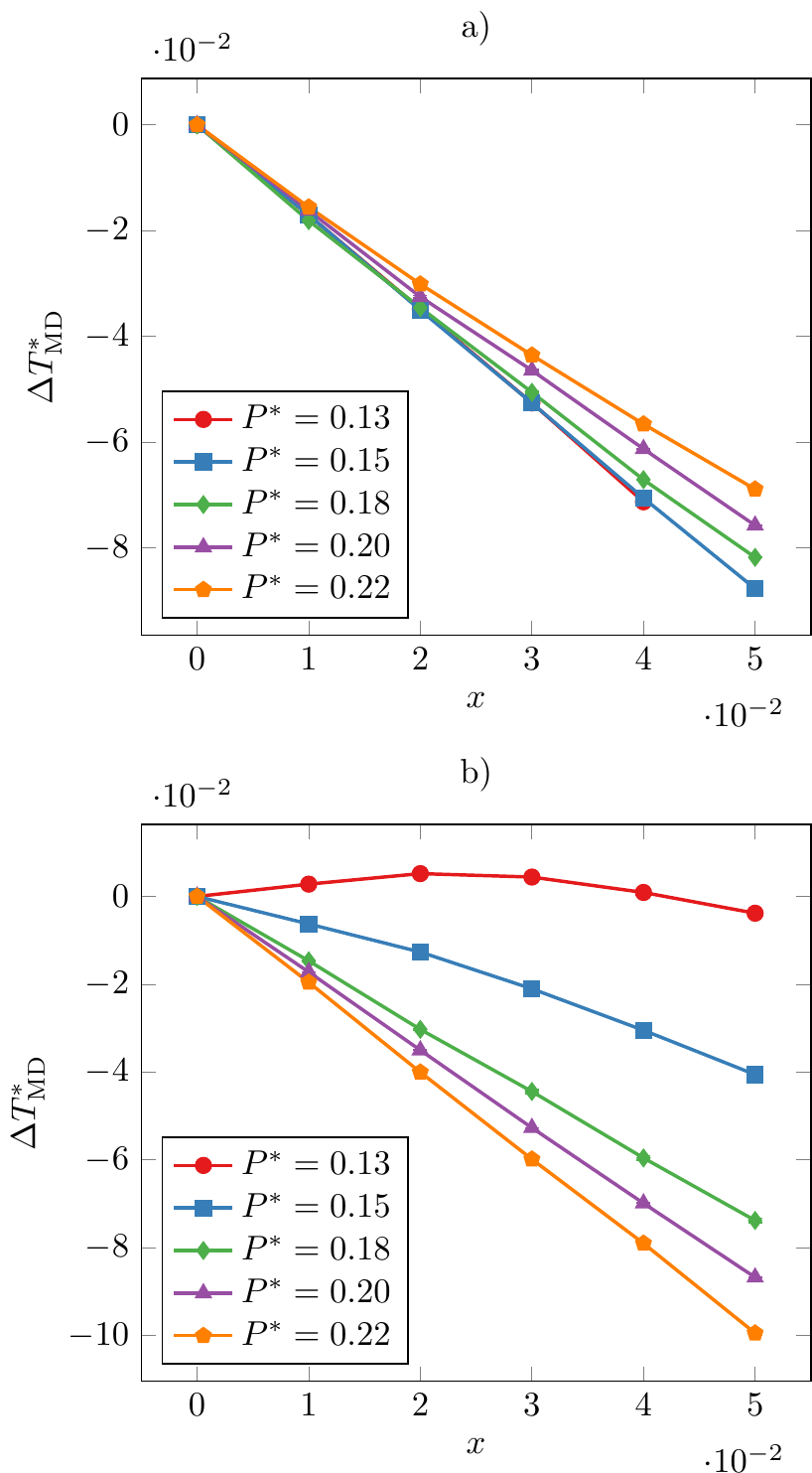}
    \caption{Change in the temperature of maximum density, $\Delta T^{*}_{\text{MD}}(x)$, as a function of solute concentration for (a) $\epsilon^{*}_{AB}=0.8$ and (b) $\epsilon^{*}_{AB}=1.1$ for reduced pressures 0.13, 0.15, 0.18, 0.20 and 0.22.}
    \label{fig:3d_dtmd}
\end{figure}
%%%%%%%%%%%%%%%%%%%%%%%% FIGURE 10 %%%%%%%%%%%%%%%%%%%%%%%%
Finally, we analyzed the impact of pressure on the enhancement of the TMD with the addition of solute. Figure~\ref{fig:3d_dtmd}(a) illustrates how the $\Delta T^{*}_{\text{MD}}(x)$ versus $x$ is impacted by pressure for $\epsilon^{*}_{AB}=0.8$. In all simulated pressures, $\Delta T^{*}_{\text{MD}}(x) < 0$, which is consistent with the single pressure system analyzed in Figure~\ref{fig:3d_change_v0AB_dtmd}.

Figure~\ref{fig:3d_dtmd}(b) shows the behavior of $\Delta T^{*}_{\text{MD}}(x)$ versus $x$ for different pressures for $\epsilon^{*}_{AB}=1.1$. Unlike the one dimensional case shown in Figure~\ref{fig:1d_dTMD_change_pressure_aAB}(b), the $\Delta T^{*}_{\text{MD}}(x)$ decreases with the increase of pressure. In this model, more pressure lowers the $\Delta T^{*}_{\text{MD}}(x)$ curve for any solute-solvent attraction, since the same effect happens in Figure~\ref{fig:3d_dtmd}(a). We also analyzed the pressure values of $P^{*}=0.10$, $P^{*}=0.25$ and $P^{*}=0.28$, but there was no TMD for the pure system in these cases, hence they weren't included in Figure~\ref{fig:3d_dtmd}.

In real water-alcohol mixture the increase of pressure increases $\Delta T^{*}_{\text{MD}}(x)$, as shown in recent experiments~\cite{lomba2022-experiment}. Our model in three dimensions is unable to reproduce this behavior, possibly due to the smoothness of the shoulder-like length scale. 

%%%%%%%%%%%%%%%%%%%%%%%% CONCLUSIONS%%%%%%%%%%%%%%%%%%%%%%%%
\section{Conclusions}
\label{sec:conclusion}
%%%%%%%%%%%%%%%%%%%%%%%% CONCLUSIONS %%%%%%%%%%%%%%%%%%%%%%%%

In this work try to understand how the mixture of alcohol in water increases the temperature of maximum density while the mixture of water with other type of solvents decreases the TMD. Our assumption is that the mechanism behind the higher TMD is the solute-solvent attraction. In order to test our hypothesis we analyzed a mixture of two types of particles. Representing the solvent, we selected a two length scale potential which, in the absence of solute, presents a TMD line and other properties of water-like systems. The solute-solvent interaction was modeled by an attractive potential. By making this interaction more attractive we aimed to represent with the same simple model both structure breakers and structure makers systems.

Our results both in one and three dimensions support our assumption that a more attractive solute-solvent interaction leads to $\Delta T^{*}_{\text{MD}}(x)>0$ while a less attractive shows $\Delta T^{*}_{\text{MD}}(x)<0$. Comparing the two cases we observe that for large solute-solvent attraction the solute appears as isolated particles surrounded by solvent particles, forming a cage which is similar to the clathrates observed in water. This result is consistent with atomistic simulations and experiments~\cite{lomba2022-experiment}. In three dimensions, the behavior of $\Delta T^{*}_{\text{MD}}(x)$ with pressure is not consistent with the one dimensional model or with the experimental results, possibly due to the softness of our selected three dimensional potential, which suggest future investigations.
%%%%%%%%%%%%%%%%%%%%%%%%%%%%%%%%%%%%%%%%%%%%%%%%%%%%%%%%%%%%%%%%
\section{Acknowledgments}
%%%%%%%%%%%%%%%%%%%%%%%%%%%%%%%%%%%%%%%%%%%%%%%%%%%%%%%%%%%%%%%%
We thank Dr. Eduardo Rizzatti for the fruitful discussions, CAPES and CNPq agencies for the graduate fellowship and INCT-FCx for the support. We also thank the reviewers for their suggestions.
%%%%%%%%%%%%%%%%%%%%%%%%%%%%%%%%%%%%%%%%%%%%%%%%%%%%%%%%%%%%%%%%
\bibliographystyle{apsrev4-2}
\bibliography{bibliography}
%%%%%%%%%%%%%%%%%%%%%%%%%%%%%%%%%%%%%%%%%%%%%%%%%%%%%%%%%%%%%%%%

%%%%%%%%%%%%%%%%%%%%%%%%%%%%%%%%%%%%%%%%%%%%%%%%%%%%%%%%%%%%%%%%
\appendix
%%%%%%%%%%%%%%%%%%%%%%%%%%%%%%%%%%%%%%%%%%%%%%%%%%%%%%%%%%%%%%%%
\section{Partition Function}
\label{sec:partitionfunction:appendix}
%%%%%%%%%%%%%%%%%%%%%%%%%%%%%%%%%%%%%%%%%%%%%%%%%%%%%%%%%%%%%%%%
In order to account for the number of different configurations for two types of particles, $A$ and $B$, we begin with a discrete model following the ideas of Ref.~\cite{prigogine1957molecular}. Consider a line of sites separated by a distance $\eta$. Each site could be occupied by a particle or remain empty. Let $N_A$ and $N_B$ represent the number of particles of each type. We denote $N$ the total number of particles and L the total size of the system. Consider two neighbor particle of type $i$ and $j$. The distance between them can be expressed with an integer $k$ as $k\eta$. We call $V^k_{ij} = V_{ij} (k\eta)$ the potential of one over the other. Let $\nu^k_{ij}$ be number of first neighbor interactions between particles of types $i$ and $j$ at a distance $k\eta$.

The total number of interactions $N_{ij}$ between particles of type $i$ and particles $j$ can be written summing over all distances $k\eta$ as
\begin{equation}
    N_{ij} = \sum_k \nu^k_{ij}.
    \label{eq:nu_kij_discrete}
\end{equation}
Since $N-1\approx N$,
\begin{equation}
    \sum_k (\nu^k_{AA}+\nu^k_{AB}+\nu^k_{BB}) = N_{AA} + N_{AB} + N_{BB} \approx N.
    \label{eq:boundary_condition_1}
\end{equation}
The length $L$ can also be expressed in terms of $\nu^k_{ij}$:
\begin{equation}
    \sum_k (\nu^k_{AA}+\nu^k_{AB}+\nu^k_{BB})k\eta = L.
    \label{eq:boundary_condition_2}
\end{equation}
Finally, the number of particles and the number of interactions is related by
\begin{equation}
    \sum_k (2\nu^k_{ii}+\nu^k_{ij}) = 2N_i.
    \label{eq:boundary_condition_3}
\end{equation}

The configurational term (that is, without the momentum) of the Canonical Partition Function is

\begin{widetext}
    \begin{equation}
        Q = \sum_{\{\nu^k_{ij}\}}
        \frac{N_A!N_B!}{\prod_k \nu^k_{AA}! \nu^k_{BB}! \left[\left(\frac{\nu^k_{AB}}{2}\right)!\right]^2} \exp{\left(-\beta\sum_k \nu^k_{AA}V^k_{AA}+\nu^k_{AB}V^k_{AB}+\nu^k_{BB}V^k_{BB}\right)},
    \end{equation}
\end{widetext}
where the term outside the exponential accounts for the different configurations with the same energy.

We multiply $Q$ by a factor $e^{-\beta P L}$ and sum over all volumes to find the partition function in the Isothermal-Isobaric ensemble. We use~\ref{eq:boundary_condition_2} to replace $L$ and, as typical in Statistical Mechanics, approximate the sum by its largest term. To find the equilibrium values of $\nu^k_{ij}$, $\ln(Y^{*})$ must be extremized. To satisfy the constraints defined by~\ref{eq:boundary_condition_3}, we introduce the Lagrange Multipliers $\lambda_1$ and $\lambda_2$. We use the Stirling approximation, remove the constants and group in terms of each $\nu^k_{ij}$. Hence, it can be seen that this extremization is satisfied by
\begin{align}
    \nu^k_{AA} &= \exp{[-\beta (V^k_{AA}+Pk\eta) +2\lambda_1]} \label{eq:max_aa}, \\
    \nu^k_{BB} &= \exp{[-\beta (V^k_{BB}+Pk\eta) +2\lambda_2]} \label{eq:max_bb}, \\
    \nu^k_{AB} &= \exp{[-\beta (V^k_{AB}+Pk\eta) +\lambda_1 + \lambda_2]}.\label{eq:max_ab}
\end{align}
Hence we find that
\begin{align*}
    \ln Y^{*} = N &+ \ln(N_A!) + \ln(N_B!) - 2\lambda_1 N_{AA} \\
    &- 2\lambda_2 N_{BB} - (\lambda_1 + \lambda_2) N_{AB}.
\end{align*}

We can write $N_{AA}$ as $N_{AA}=e^{2\lambda_1}\varphi_{AA}$, where we define
\begin{equation}
    \varphi_{ij} \equiv \sum_k e^{-\beta (V^k_{ij}+Pk\eta)}.
\end{equation}
Using the analogous relations for $N_{AB}$ and $N_{BB}$ we conclude that
\begin{align*}
    2\lambda_1 &= \ln\left(\frac{N_{AA}}{\varphi_{AA}}\right), \\
    2\lambda_2 &= \ln\left(\frac{N_{BB}}{\varphi_{BB}}\right), \\
    \lambda_1+\lambda_2 &= \ln\left(\frac{N_{AB}}{\varphi_{AB}}\right).
\end{align*}
Replacing the $\lambda$'s in $\ln Y^{*}$ with these expressions and exponentiating both sides of the equation leads to the expression given in Section~\ref{sec:model}, with the $\Lambda$ term coming from the momentum integration. In the limit of $\eta \rightarrow 0$, we must replace the sum in $\varphi_{ij}$ by an integral:
\begin{equation}
    \varphi_{ij} = \int_{0}^{\infty} e^{-\beta(V_{ij}(r)+Pr)}dr.
\end{equation}
%%%%%%%%%%%%%%%%%%%%%%%%%%%%%%%%%%%%%%%%%%%%%%%%%%%%%%%%%%%%%%%%
\section{Exact Gibbs Free Energy}
\label{sec:gibbs:appendix}
%%%%%%%%%%%%%%%%%%%%%%%%%%%%%%%%%%%%%%%%%%%%%%%%%%%%%%%%%%%%%%%%
The Gibbs Free Energy is
\begin{widetext}
    \begin{align}
        g(\beta, P) = &-\frac{1}{\beta}\lim_{N \rightarrow \infty} \left[ \frac{\ln Y(\beta, P, N)}{N} \right] \\
        =&\frac{1}{\beta} \ln \Lambda -\frac{1}{\beta}\lim_{N \rightarrow \infty} \left[\frac{\Delta}{N} + \frac{N_{AA} \ln \varphi_{AA} + N_{AB} \ln \varphi_{AB} + N_{BB} \ln \varphi_{BB}}{N} \right]
        \label{eq:gibbs_free_energy},
    \end{align}
\end{widetext}
where
\begin{equation}
    \Delta \equiv \ln \left[\frac{N_A! N_B!}{N_{AA}!N_{BB}! \left[\left(\frac{N_{AB}}{2}\right)!\right]^2}\right]
\end{equation}
In the continuous limit, we can write~\ref{eq:nu_kij_discrete} as
\begin{equation}
    N_{ij} = \int_{0}^{\infty} \nu_{ij}(r) dr.
    \label{eq:nu_kij_continuous}
\end{equation}
where the $\nu_{ij}(r)$ are the terms can be found from the minimization process. Replacing this in~\ref{eq:boundary_condition_3} results in coupled quadratic equations. This can be solved for $e^{2\lambda_1}$ and $e^{2\lambda_2}$.

From~\ref{sec:partitionfunction:appendix}, we know that
\begin{align}
    N_{AA} &= e^{2\lambda_1}\varphi_{AA},\\
    N_{BB} &= e^{2\lambda_2}\varphi_{BB}.
\end{align}
Dividing by $N$ and using the expressions found for $e^{2\lambda_1}$ and $e^{2\lambda_2}$:

\begin{widetext}
    \begin{align}
        x_{AA} &= \lim_{N \rightarrow \infty} \frac{N_{AA}}{N} = \frac{(x_A-x_B) - 2x_A\gamma + \sqrt{(x_A-x_B)^2 + 4x_Ax_B\gamma}}{2\left(1 - \gamma\right)},\\
        x_{BB} &= \lim_{N \rightarrow \infty} \frac{N_{BB}}{N} = \frac{-(x_A-x_B) - 2x_B\gamma + \sqrt{(x_A-x_B)^2 + 4x_Ax_B\gamma}}{2\left(1 - \gamma\right)},\\
        x_{AB} &= \lim_{N \rightarrow \infty} \frac{N_{AB}}{N} = 2(x_A-x_{AA}),
    \end{align}
\end{widetext}
where we used used $N_{AB} = 2(N_A - N_{AA})$ from~\ref{eq:boundary_condition_3} and $\gamma \equiv \varphi_{AA}\varphi_{BB}/\varphi_{AB}^2$. The result of~\ref{sec:model} is found by replacing these expressions in~\ref{eq:gibbs_free_energy}.

\end{document}